\documentclass[prl,twocolumn, showpacs, amsmath]{revtex4}

\usepackage{amssymb}
\usepackage{amstext}
\usepackage{amscd}
\usepackage{amsthm}
\usepackage{amsopn}
\usepackage{amsmath}
\usepackage{amsbsy}
\usepackage{indentfirst}
\usepackage{graphicx}

\newcommand{\be}{\begin{equation}}
\newcommand{\ee}{\end{equation}}
\newcommand{\bea}{\begin{eqnarray}}
\newcommand{\eea}{\end{eqnarray}}

\def\<{{\langle}}
\def\>{{\rangle}}

\begin{document}

\author{Pankaj Mehta }
\affiliation{ Dept. of Molecular Biology and Dept. of Physics, Princeton University, Princeton, NJ 08544}
\author{Natan Andrei}
\affiliation{Center for Materials Theory, Rutgers University,
Piscataway, NJ 08854}

 \title{ Nonequilibrium quantum-impurities: from entropy production to information theory  }

\begin{abstract}

Nonequilibrium steady-state currents, unlike their equilibrium
counterparts, continuously dissipate energy into their physical
surroundings leading to  entropy production and time-reversal
symmetry breaking. This letter discusses these issues in the context of
quantum impurity models driven out of equilibrium by attaching the
impurity to leads at different chemical potentials and temperatures.
We start by pointing out that entropy production is often hidden in
traditional treatments of quantum-impurity models. We then use
simple thermodynamic arguments to define the rate of entropy
production, $\sigma $.  Using the scattering framework recently
developed by the authors we show that $\sigma$ has a simple
information theoretic interpretation in terms of the Shannon entropy
and Kullback-Leibler divergence of nonequilibrium distribution
function.  This allows us to show that the entropy production is
{\it strictly positive} for any nonequilibrium steady-state.  We conclude
by applying these ideas to the Resonance Level Model and the Kondo
model.
\end{abstract}

\pacs{72.63.Kv, 72.15.Qm, 72.10.Fk   }

\maketitle

Over the last decade, it has become possible to experimentally explore
quantum impurity models in nonequilibrium settings \cite{GG}.  Quantum
impurities are most often realized using quantum dots, small pools of
electrons confined in space. In a typical experimental set-up, a
quantum dot is attached to multiple leads at different chemical
potentials resulting in a nonequilibrium current across the dot.  This
opens up the exciting possibility of using quantum impurity models to
experimentally and theoretically explore nonequilibrium physics.

We focus in this paper on the case when the currents across the dot
are in a nonequilibrium steady-state (NESS).  NESS's have been
extensively studied in a variety of physical systems including
lattice models and fluid systems \cite{Reichl}.  A system in a NESS continuously dissipates
energy into its surrounding, resulting in a continuous production of
entropy and the breaking of time-reversal symmetry. This is in
contrast with equilibrium steady-states where a persistent current
can flow without producing entropy or dissipating energy.

NESS's in quantum-impurity systems are usually modeled by
adiabatically turning-on the interaction between the impurity and the
leads in the far past, at a time $t_0 <0 $, and then evolving the
system in time to the present, $t=0$.  Thus at $t < t_o$ the system is
described by a density matrix $\rho_o$ describing two non interacting
leads (reservoirs) at different chemical potentials $\mu_i$ and
different temperatures $T_i$ and the uncoupled impurity. At time
$t=t_0$ the impurity is coupled to the leads and evolves adiabatically
according to $H= H_o + \theta(t-t_0) e^{\eta t} H_1$, i.e. at a time
$t$ it is described by a time evolved density operator $\rho(t)
=U^{\dagger}(t, t_0)\rho_o U(t, t_0)$, where the evolution operator
$U(t, t_0)$ corresponds to $H(t)$.  The non-equilibrium density matrix
is used to compute non-equilibrium expectation values $\langle \hat{O}
\rangle = Tr {\rho(t) \hat{O}}$. This expectation value becomes time
independent and a steady-states emerges if the leads are good thermal
baths and the system is {\it open}. Namely, both the number of
particles in the lead, $N_i$, and the size of the lead, $L$, are
infinite, with the limit taken ab initio. The establishment of a
steady state follows, in this language, from the existence of the
open-system limit $1/L \ll 1/ |t_o| \ll \eta \to 0$. In this case, the
density matrix becomes time-independent \cite{Doyon}. We denoted it by
$\rho_s$.  Most treatments of the problem are based on calculating
$\rho_s$ in various ways. A large class of theoretical treatments use
Keldysh perturbation theory to calculate $\rho_s$ \cite{Doyon,
Keldysh}, while others use TBA-based approaches \cite{BS}. The
Y-operator treatments calculate $\rho_s$ using a recursive algorithm
\cite{Yop}.  The scattering framework recently proposed by the authors
uses scattering theory to calculate $\rho_s$ in a time-independent
manner \cite{Mehta}.  What is common to all these treatments is that
the time-independent nonequilibrium density matrix, $\rho_s$, commutes
with the Hamiltonian of the system, $H$.  This observation raises the
intriguing question of how a density matrix that commutes with the
Hamiltonian capture the defining characteristics of nonequilibrium
physics, such as energy and particle currents, energy dissipation and
entropy production.

 In an {\it open} system it is not necessary to include explicit
 mechanisms, such as phonons, that allow relaxation of the high-energy
 electrons transferred between leads. Instead, energy dissipation can
 be effectuated by implementing the open-system limit. Since the size
 of the system is much larger than the turn-on time, the high-energy
 electrons transferred between leads continue with the same energy off
 to edges of the leads and "dissipate" their energy infinitely far
 away. The limit also ensures that once the electrons cross the
 impurity they can not return to the system, giving rise to
 time-reversal symmetry breaking. In the language of Green's
 functions, the open-system limit induces the poles in the self-energy
 to merge into a branch cut, leading to dissipative effects
 \cite{Doniach}. Under these circumstances the nonequilibrium physics
 is captured by {\it scattering eigenstates} -- eigenstates of the
 Hamiltonian $H_o +H_1$ defined on the open system with appropriate
 asymptotic boundary conditions \cite{Mehta}. Their existence is
 follows from the open-system limit by the Gellman-Low theorem
 \cite{GellMann}

In the  time-independent pictured that emerges after taking the
appropriate limits, time has been traded for space
\cite{Merzbacher}. It is convenient to follow the standard steps to
express the the impurity problem in a one dimensional language
 and "unfold" the leads \cite{alj}.  One then has only
right movers, with the incoming particles in the region $x<0$ and
the outgoing particles in the region $x>0$ and the impurity at
$x=0$. In this picture, the far past corresponds to the region $x
\ll 0$ and the far future to the region $x \gg 0$. These can be
neatly described by introducing the asymptotic 
 incoming and outgoing single-particle distribution
functions \cite{Mehta2, note}. 
 Denote the distribution function of the
incoming particles (in the region $x\ll 0$) in lead $i$ by
$f_i^-(p_\alpha)$.  After scattering off the impurity at $x=0$, the
outgoing particles in lead $i$ are also asymptotically described ($x
\to \infty$) by a distribution function for the outgoing particles
in lead $i$, $f^{+}(p_\alpha)$.  The incoming particles from each
lead can be described asymptotically by the Fermi function
$f_{FD}(p_\alpha)$ at the temperature and chemical potential
appropriate to each lead.  However, due to the scattering at the
leads, $f_i^{+}(p_\alpha)$ is no longer the Fermi distribution
function but instead a nonequilibrium distribution function that
encodes information about scattering. The distribution
functions, $f_i^{\pm}(p_\alpha)$, contain all the information about
the nonequilibrium energy and particle currents 
since these are computed from local single-particle operators.

\begin{figure}
\includegraphics[width=\linewidth]{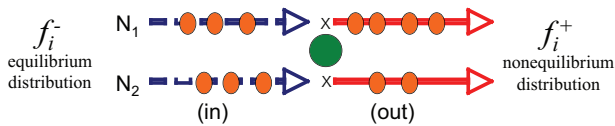}
\caption{Incoming and outgoing distribution functions}
 \vspace{-0.5cm}
\end{figure}

We now discuss the entropy production in the system. We begin with
thermodynamic considerations, then reformulated the problem in terms
of non-equilibrium distributions which emerge naturally in the
scattering formalism. We then relate our expressions for entropy
production to information theoretic quantities of the underlying
nonequilibrium distribution functions. This latter formulation allows
us to show that the entropy production is due to two underlying
process, {\it mixing} and {\it relaxation} for which we give explicit
expressions. Finally we prove that the rate entropy production is
 strictly positive in the NESS.
As an
application of these concepts, we explore how strong-correlations
manifest themselves at the level of entropy production in the Kondo
model.

The thermodynamic definition of the rate of entropy production
follows naturally from the observation that a quantum-impurity
coupled to leads is a discontinuous system: the two reservoirs are
connected to each other by a single impurity \cite{DeGroot}.  In
such a system, all entropy is produced in the leads and the entropy
produced at the quantum-impurity itself is negligible. Recall that
the differential entropy of a system, $dS$, is related to $\delta Q$
the heat that flows into or out of the system via $TdS= \delta Q$
with $T$ the temperature of the system. For a discontinuous system
with two leads, this allows us define the rate of entropy
production, $\sigma$, as \be \sigma \equiv \frac{dS}{dt} \equiv \frac{1}{T_1}
\frac{\delta Q_1}{dt} +\frac{1}{T_2}\frac{\delta Q_2}{dt}
\label{def1sigma} \ee where $\delta Q_i$ is the heat that flows into
lead $i$, and $T_i$ is the temperature of lead $i$.  The heat produced
by a lead is related to the change in energy and particle number in
the lead by $\delta Q_i = dE_i -\mu_i dN_i$. Furthermore, since
neither particles nor energy cannot disappear at the quantum impurity
in a nonequilibrium steady-state we have the conservation laws $
\frac{dN_1}{dt}= -\frac{dN_2}{dt}$ and $ \frac{dE_1}{dt}=
-\frac{dE_2}{dt}$, and therefore, $\sigma = \left(
\frac{1}{T_1}-\frac{1}{T_2}\right)\frac{dE_1}{dt}
-\left(\frac{\mu_1}{T_1}-\frac{\mu_2}{T_2}\right)\frac{dN_1}{dt}
\label{defsigma} $. Note that an important ingredient in our
 consideration has been the inclusion of implicit relaxation
mechanisms (e.g.  the open-system limit or phonons) in each lead that
equilibriate high energy electrons.  This allows us to characterize
the lead $i$ by a temperature $T_i$ even in the presence of
nonequilibrium currents.  As a result, we shall see that the entropy
production contains not only a mixing term but also a term that takes
into account the effect of relaxation processes.

This expression can be related to the nonequilibrium currents across
the dot by noting that in a NESS, the rate of change in energy of lead
1 is the expectation value of the energy current across the dot,
$\frac{dE_1}{dt}= \<I_E\>_s$ and the rate of change in particle number
in lead 1 is the expectation value of the nonequilbirum particle
current, $\frac{dN_1}{dt}= \< I_N \>$. Thus, in terms of the currents
across the dot, the rate of entropy production takes the form
\be
 \sigma = -\left(\frac{1}{T_1}-\frac{1}{T_2}\right)\<I_E\>_s
+\left(\frac{\mu_1}{T_1}-\frac{\mu_2}{T_2}\right)\<I_N\>_s
\label{entropycurrent}
\ee
For the special case, when the temperatures of the two leads are
equal, $T_1=T_2=T$, one has $ \sigma_{T} = \frac{V
\<I\>_s}{T}$ where we have defined the voltage $V=\mu_1-\mu_2$. The
term in the numerator of the last expression is  the familiar
power of an electrical circuit, $P=\<I\>_s V$.  Thus, for any
quantum-impurity model where the two leads are at the same
temperature, the rate of entropy production has the simple
interpretation as the power across the circuit divided by the
temperature.

We now show how to define $\sigma$ directly in terms of information
theoretic quantities involving nonequilibrium distribution functions
$f^{\pm}_i$ introduced above.  We first express the energy and
particle currents in terms of $f_i^{\pm}$.  In a steady-state the
particle current $\<I \>_N$ can be calculated in the time-independent
picture using the relation $\< I_N \> = \frac{dN_1}{dt}= \lim_{L
\rightarrow \infty} \frac{N_1(L/2) -N_1(-L/2)}{L/v_F}$ with $N_1(\pm L/2)=
\sum_{ p_\alpha} f_1^\pm(p_\alpha)$ the asymptotic particle number of
the outgoing and incoming electrons and $v_F$ is the Fermi velocity
\cite{note3}. In writing this
expression, we have used the fact that the leads are non-interacting
and that in the time-independent picture the far past corresponds to
$x \ll 0$ and the far future to $ x \gg 0$. One can also define the
energy current in an analogous manner by the expression $\< I_E \> =
\frac{dE_1}{dt}= \lim_{L \rightarrow \infty} \frac{E_1(L/2)
-E_1(-L/2)}{L/v_F}$ where $E(\pm L/2)= \sum_{p_\alpha}
\epsilon_{p_\alpha}f_i^\pm(p_\alpha) $ measures the asymptotic energy
of the incoming and outgoing particles and $\epsilon_{p_\alpha}$ is
the energy dispersion for the free electrons. Substituting these
expressions into (\ref{entropycurrent}) and using the conservation
laws relating the energy and particle number in the two leads, one can
write the rate of entropy production in terms of these distribution
function as,
 \be \sigma = \lim_{L \to
\infty} \frac{1}{L}\sum_{\alpha} \sum_{i=1,2}v_F [f_i^+(p_\alpha) - f_i^-(p_\alpha)]
\frac{\epsilon_{p_\alpha} - \mu_\alpha}{T_i}.
\label{temp}
\ee

 We proceed to interpret our results in information-theoretic terms and
  rewrite $\sigma$ in terms of the Shannon entropy current
 and the Kullback-Leibler distance of the probability
distributions $f_{i}^\pm$. Doing so will allow us to identify to
mechanism underlying entropy production and prove that it is
strictly positive. The Shannon entropy, $S_{\rm IT}[q(x)]$ of a
probability distribution $q(x)$ measures the uncertainty of a random
variable $X$ that takes discrete values in the set $\{x_\beta\}$
with probability $q(x)$ and is defined as $ S_{\rm IT}[q] \equiv
-\sum_{\beta} q(x_\beta) \ln{q(x_\beta)}$ \cite{Jaynes, Shannon}.
The relative entropy or Kullback-Leibler (KL) distance between the
distribution $q(x)$ and $s(x)$, \be D_{\rm KL}[s(x)||q(x)]=
\sum_{x_\beta} s(x_\beta) \ln{\frac{s(x_\beta)}{q(x_\beta)}}
\label{defKL} \ee measures the inefficiency in assuming that the
distribution is
 $q(x)$ when the true distribution is $s(x)$ \cite{Shannon}.

The entropy of a free electron bath, usually defined
thermodynamically using the partition function,
 can also be defined by means of
information theoretic ideas as the Shannon entropy of the Fermi-Dirac
distribution function $f_{FD}(p_\alpha)$,
\bea
S_{\rm IT}[f_{FD}] &=& -\sum_\alpha(1- f_{FD}(p_\alpha))
\ln{(1-f_{FD}(p_\alpha)) } \nonumber \\ &-& \sum_{\alpha}
f_{FD}(p_\alpha) \ln{f_{FD}(p_\alpha)}
\label{entropydef}
\eea
A short calculation shows that this definition is equivalent to the
usual thermodynamic definition of entropy in for free electrons in
thermal and chemical equilibrium \cite{Kubo}.  The advantage of the
Shannon entropy as compared to thermodynamic entropy is that it has
a natural generalization to nonequilibrium systems, namely the
Shannon entropy of the nonequilibrium distribution function.

To rewrite the rate of entropy production in terms of the information
theoretic quantities defined above we make use of the fact that the
incoming particles from lead $i$ are distributed according the
Fermi-Dirac distribution function, $f_i^{-}(p_\alpha)=
f_{iFD}(p_\alpha)$. As a result, we know that we can write
$(\epsilon_p-\mu_i)/T_i= \ln{[(1- f_i^-(p_\alpha))/f_i^-(p_\alpha)]}
$. Substituting this expression into (\ref{temp}) and using the
definitions (\ref{entropydef}) and (\ref{defKL}) one gets
\bea \sigma &=& \lim_{L \to \infty} \frac{1}{L} \sum_{i} v_F[S_{\rm
IT}[f_i^+] -S_{\rm IT}[f_i^{-}] ] + v_F D_{\rm KL}[f_i^ + || f_i^-]
\nonumber \\ &\equiv& \sum_{i} \Delta \sigma_i + v_F \lim_{L \to
\infty} \frac{1}{L} \sum_{i} D_{\rm KL}[f_i^+ ||f_{iFD}],
\label{defITentropy}
\eea
where in the second line we defined $\sigma_i$, the Shannon-entropy
produced per unit time at lead $i$. The only assumption we  used
 is that the leads are non-interacting. Thus, this
expression is valid for {\it all} nonequilibrium quantum-impurity
models with non-interacting leads where no entropy is produced at the
impurity.

The two terms in (\ref{defITentropy}) have a clear physical
meanings. $\Delta \sigma_i$ measures the increase in the entropy per
unit time of the system due to mixing of electron between leads. 
In information
theoretic terms, it measures the extra uncertainty about electrons
introduced because electrons are being transferred between leads.
This term does not however take into account the extra entropy
produced in the physical system by the relaxation to equilibrium of
high energy electrons transferred between leads.  This is captured
by the second term in (\ref{defITentropy}), the KL-distance between
the nonequilibrium distribution function of the outgoing electrons
of lead $i$ and the thermal equilibrium Fermi-Dirac distribution
function. Thus, in thermodynamic terms, we can interpret $T_i D_{\rm
KL}[f_i^+||f_{iFD}]$ as the available work that can be extracted
when a system with the nonequilibrium distribution function
$f_i^+(p)$ relaxes to equilibrium at temperature $T_i$.  A similar
interpretation of the KL-distance has been suggested in the theory
of nonequilibrium chemical reaction \cite{Proccacia}. In information
theoretic terms, $D_{\rm KL}[f_i^+||f_{iFD}]$ measures the
uncertainty introduced by the relaxation of high-energy electrons.
Thus, despite the fact that, as discussed in the introduction, we
have not included  physical energy dissipation mechanisms such as
phonons in our system, information theoretic considerations still
allow us to extract the amount of entropy produced by relaxation of
high energy electrons to equilibrium.

The expression (\ref{defITentropy}) further allows us to adapt
Shannon's original proof that a communication device always
increases the entropy of a message \cite{Shannon} to prove that for
a NESS, $\sigma > 0$.  To do so, we view the scattering problem as a
{\it classical} 
communication device where a transmitter at $x \rightarrow -\infty$
prepares a message -- in this case the incoming particles in each
lead that are to be scattered -- which is then transmitted down the
communication channel -- in this case the scattering of the incoming
electrons off the impurity resulting into outgoing particles. The
receiver, at $x \rightarrow \infty$, then receives the output of
this message -- the outgoing electrons. Note that the messages
sent by our classical device
utilize only the distribution functions and not the full many-body aspects
of the problem.
 Armed with this analogy, we
apply Shannon's proof to conclude that the first term in
(\ref{defITentropy}) is greater than equal to zero, $\Delta S \ge
0$, with equality if and only if $f_i^+=f_i^-$. Furthermore, by the
definition of the KL-distance, we know that $ D_{\rm
KL}[f_i^+||f_{iFD}] \ge 0$ with equality if and only if $f_i^+ =
f_{iFD}$ \cite{Shannon}. Thus we have proven
 that for a NESS, $ \sigma >0$.

As an application, we calculate the rate of entropy production for
the Resonance Level Model (RLM), which describes a local level,
$d^\dagger$, attached to two leads of
spin-less electrons at different chemical potentials. The Hamiltonian, $
 H_{RLM} =
\sum_{i=1,2} \left[ \int \psi_{i}^\dagger(x) \partial_x
\psi_{i}(x) + t(d^\dagger \psi_{i}(0) + h.c) +\epsilon_d
d^\dagger d \right]$,
 is quadratic and the entropy production rate can be easily
 computed from the thermodynamic definiton (\ref{entropycurrent})
\bea
\sigma= \frac{1}{T_1 T_2}\int
\frac{dp}{2\pi} \left[(p-\mu_2)(T_1-T_2) +V T_2 \right] \nonumber \\
\times \left[ \frac{\Gamma^2}{(p-\epsilon_d)^2 + \Gamma^2} [f_{1FD}
(p)-f_{2FD}(p)]\right].
\label{RLMentropy}
\eea
Alternatively,  we can compute the entropy using the nonequilibrium scattering distribution
functions $f_{i}^\pm$ calculated using the scattering method  and explicitly given by
\be
f_{1,2}^+ = f_{1,2\,FD}(p) |R(p)|^2 + f_{2,1 \,FD}|T(p)|^2
\label{distr}
\ee
with $| T(p)|^2= \frac{\Gamma^2}{\Gamma^2 +
(p-\epsilon_d)^2} $ the probability that a particle with the energy $p$
is transmitted across leads and $|R(p)|^2=1-|T(p)|^2$ the
probability that a particle stays in the same lead. Plugging into (\ref{defITentropy}) and simplifying 
we again find that $\sigma$ is of the form (\ref{RLMentropy}).

As a final application of these ideas, we explore how the Kondo effect
manifests itself at the level of entropy production.  We consider the
experimentally relevant case where the temperatures of the two leads
are identical. The particle current in the Kondo model at $ V \gg T_k$
is found to be $\<I\>_s \sim V/ \log^2{\frac{\sqrt{V^2+T^2}}{T_k}}$
\cite{Keldysh, Doyon}.  Thus, at high voltages $ \sigma \sim V^2
/\log^2{\frac{\sqrt{V^2+T^2}}{T_k}}$.  At very small voltages
 the problem can be treated by linear response around the
strong-coupling fixed point. In the regime, $T \ll V \ll T_k$, the
current is given by \cite{Glazman}, $ \<I\>_s \sim V
-\frac{3V^3}{2T_k^2}$ and the entropy production is just $\sigma
\equiv V^2/T (1- \frac{3V^2}{2T_k^2})$. We can compare this at low
temperatures to the RLM -- which describes the equilibrium  strong
coupling Kondo physics -- where $\sigma \sim V^2/T(1- \frac{V^2}{8\pi
\Gamma^2})$ (taking $\mu_{1,2}= \epsilon_d \pm V/2$).  Thus, as
expected, from the point of view of entropy production, the Kondo
model behaves essentially as a local level with hybridization
proportional to the Kondo temperature at low voltages and
temperatures. At high temperature the correlations weaken, leading to
an increase in the rate of entropy production compared with the RLM.

Nonequilibrium physics remains relatively poorly understood compared
to its equilibrium counterpart.  It is our belief that thinking about
the nonequilibrium physics of quantum-impurity models is likely to
yield general concepts applicable to a wide variety of physical
systems. In this letter, we have shown how entropy production in a
NESS has a natural interpretation in terms of information theoretic
quantities involving the nonequilibrium distribution function. We
believe that this is  the first step towards a broadening our
understanding of the rich nonequilibirum phenomenon present in these
models. In particular, in light of the close connections between
quantum information theory and quantum-entanglement, it would be
particularly interesting to understand if the rate of entropy
production has a simple interpretation in terms of entanglement
entropies.

{\bf Acknowledgments}: We would like to thank Eran Lebanon, Joel
Lebowitz, Achim Rosch, Alessandro Silva, and Emil Yuzbashyan for
discussions and comments on the text, and especially Sasha
Zamolochikov for posing the questions that lead us to consider the
topics in this paper. This work was partially supported by NSF grant
4-21776.


\begin{thebibliography}{999}
\bibitem{note} As is usual in scattering theory, we quantize the
momenta $\{ p_{\alpha} \}$ by placing the system by in a box of length
$L$ and then taking the limit $L \to \infty$.
\bibitem{note3} We follow standard practice, valid when temperatures
and voltages are small compared to $E_F$, of linearizing the spectrum
around the Fermi surface so that all particles move with same Fermi
velocity.
\end{thebibliography}
\end{document}